\documentclass[pdflatex,sn-mathphys-num,iicol]{sn-jnl}

\usepackage{graphicx}%
\usepackage{multirow}%
\usepackage{amsmath,amssymb,amsfonts}%
\usepackage{amsthm}%
\usepackage{mathrsfs}%
\usepackage[title]{appendix}%
\usepackage{xcolor}%
\usepackage{textcomp}%
\usepackage{manyfoot}%
\usepackage{booktabs}%
\usepackage{algorithm}%
\usepackage{algorithmicx}%
\usepackage{algpseudocode}%
\usepackage{listings}%

\theoremstyle{thmstyleone}%
\theoremstyle{thmstyletwo}%

\theoremstyle{thmstylethree}%

\newcommand{\ii}{\mathrm{i}} 

\newcommand{\tomega}{\tilde{\omega}}

\newcommand{\vk}{ \mathbf{k} }

\newcommand{\imp}{ \mathrm{imp} }
\newcommand{\SC}{ \mathrm{SC} }
\newcommand{\kB}{ k_{\mathrm{B}} }

\newcommand{\Tr}{ \mathrm{Tr} }
\newcommand{\Qan}{Q_{a n}}
\newcommand{\Qbn}{Q_{b n}}
\newcommand{\Qalphan}{Q_{\alpha n}} 
\newcommand{\Qbetan}{Q_{\beta n'}} 
\newcommand{\Dimp}{D}

\begin{document}

\title[Grand thermodynamic potential in unconventional superconductor]{Grand thermodynamic potential in a two-band unconventional superconductor}


\author*[1]{\fnm{Vadim A.} \sur{Shestakov}}\email{v\_shestakov@iph.krasn.ru}
\author[1]{\fnm{Maxim M.} \sur{Korshunov}}

\affil*[1]{\orgname{Kirensky Institute of Physics, Federal Research Center KSC SB RAS}, \orgaddress{\city{Krasnoyarsk}, \postcode{660036}, \country{Russia}}}


\abstract{In a two-band system, both conventional sign-preserving $s_{++}$ and unconventional sign-changing $s_{\pm}$ superconducting state may appear at low temperatures. Moreover, they may transform from one to another due to the impurity scattering. To study the details of such a transition here we derive the expression for the Grand thermodynamic potential $\Omega$ for a two-band model with nonmagnetic impurities considered in a $\mathcal{T}$-matrix approximation. For the iron-based materials within the multiband Eliashberg theory, we show that the $s_{\pm} \to s_{++}$ transition in the vicinity of the Born limit is a first order phase transition.}




\maketitle

\section{Introduction}\label{sect:1.intro}
Iron-based materials represent a prototypical system for studying various nontrivial states including magnetic fluctuations and long-range order, nematic order, and unconventional superconducting state~\cite{SadovskiiReview2008,StewartReview}. It is a multiband system with the possible $s_{\pm}$ superconducting gap that changes sign between the bands~\cite{HirschfeldKorshunov2011}. The $s_{\pm}$ gap is naturally follows from the spin-fluctuation theory of Cooper pairing~\cite{HirschfeldKorshunov2011,Korshunov2014eng} and its presence was confirmed in a number of experiments~\cite{Prozorov2014,Inosov2016,Schilling2016,Korshunov2018,Ghigo2018,KorshunovKuzmichev2022}.

Nonmagnetic disorder doesn't harm conventional superconductivity. On the other hand, for the unconventional superconductor different types of behaviour take place~\cite{Golubov1997,Ohashi2004,KorshunovUFN2016}. The sign-changing order parameter may be suppressed by the impurity scattering or may transform to the sign-preserving one. The latter corresponds to the $s_{\pm} \to s_{++}$ transition~\cite{EfremovKorshunov2011,ShestakovKorshunovSymmetry2018}. As further studies have shown, the transition is not smooth at very low temperatures in the vicinity of the Born limit~\cite{ShestakovKorshunovSUST2018}. The result is based on the analysis of the gap function changes with the impurity scattering rate. The more informative approach, however, would include consideration of systems' energy and its evolution with the parameters of the model. To pursue this goal, one have to write the expression for the grand thermodynamic potential $\Omega$ that is sometimes also called a Landau free energy. While for a one-band system the result is known for a long time, the two-band system with the impurity scattering possess a challenge due to the complexity of the matrix equations. Here we show how to overcome the difficulty and derive the grand thermodynamic potential $\Omega$ for a two-band model of an unconventional superconductors with the nonmagnetic disorder treated in the $\mathcal{T}$-matrix approximation. Then we calculate the difference $\Delta\Omega$ of the thermodynamic potential for the normal ($\Omega_N$) and the superconducting ($\Omega_S$) states in the model of the $s_{\pm}$ state in iron-based materials. There are two competing solution of the Eliashberg equations in the narrow range of temperatures and impurity scattering rates $\sigma$~\cite{Shestakov_Korshunov_SUST2025,FTT2025}. The situation reminds a one with the magnetically frustrated system -- there are several competing states and the system have to `choose' one of them. With the help of $\Delta\Omega$ we are able to discern the unique solution. Since in that case $\Delta\Omega$ has a kink, we conclude that the abrupt $s_{\pm} \to s_{++}$ transition is the first order phase transition.

\section{Grand thermodynamic potential}\label{sect:1.potential}
\subsection{General treatment}
To calculate the Grand thermodynamic potential, 
we start with the Luttinger-Ward expression for a multiband system~\cite{LW1960II,Luttinger1960} generalized to the case of a superconductor with nonmagnetic impurities,
\begin{align}
	\Omega_{\mathrm{S}}(T) &= -T\sum_{\omega_n,\vk}\Tr\left[ \ln{\lbrace-\hat{\mathbf{G}}^{-1}(\vk,\ii\omega_n)\rbrace} \right. \nonumber \\
	& + \left. \hat{\mathbf{\Sigma}}(\vk,\ii\omega_n)\hat{\mathbf{G}}(\vk,\ii\omega_n) \right] + \Omega'(T), \label{eq:LW}
\end{align}
where $\omega_n = 2(n+1)\pi T$ is the Matsubara frequency, $n$ is an integer number, $T$ is temperature in units of energy (hereinafter, we employ the natural system of units with $\kB=\hbar=1$), $\Tr[...]$ is the trace over all subspaces (Nambu space, band indexes etc.), sum (integration) over momenta $\sum_{\vk}$ is taken over the whole first Brillouin zone. Here $\Omega'(T)$ is the Luttinger-Ward functional that can be separated into superconducting ($\Omega'_{\SC}(T)$) and impurity ($\Omega'_{\imp}(T)$) parts,
\begin{equation}
	\Omega'(T) = \Omega'_{\SC}(T) + \Omega'_{\imp}(T). \label{eq:LW_f}
\end{equation}
In equation~\eqref{eq:LW}, the Green's function $\hat{\mathbf{G}}$ and self-energy $\hat{\mathbf{\Sigma}}$ both are matrices in the band (denoted by {\bf{bold face}} font) and Nambu (denoted by `$\hat{\phantom{a}}$' symbol) spaces. Generally, $\hat{\mathbf{G}}$ and $\hat{\mathbf{\Sigma}}$ are non-diagonal with respect to the band indices ($\alpha, \beta$). However, for the superconductivity the main role is played by the processes with energies close to the Fermi level, where the band non-diagonal contributions are insufficient. Thus, we can consider the Green's function and the self-energy as matrices diagonal in the band space, $\lbrace\hat{\mathbf{\Sigma}}(\vk,\ii\omega_n)\rbrace_{\alpha,\beta} \equiv \lbrace\hat{\mathbf{\Sigma}}(\vk,\ii\omega_n)\rbrace_{\alpha,\alpha} \equiv \hat{\Sigma}_{\alpha}(\vk,\ii\omega_n)$. The self-energy matrix $\hat{\mathbf{\Sigma}}(\vk,\ii\omega_n)$ has the form
\begin{align}
	\hat{\Sigma}_{\alpha}(\vk,\ii\omega_n) &= \ii\omega_n[1-Z_\alpha(\vk,\ii\omega_n)]\hat{\tau}_0 \nonumber \\
	&+ \chi_\alpha(\vk,\ii\omega_n)\hat{\tau}_3 + \phi_{2\alpha}(\vk,\ii\omega_n)\hat{\tau}_2, \label{eq:SelfEn}
\end{align}
where $Z_\alpha(\vk,\ii\omega_n)$ and $\chi_\alpha(\vk,\ii\omega_n)$ are functions renormalizing the Matsubara frequencies and quasiparticle dispersion for $\alpha$-th band, respectively, $\phi_{2\alpha }(\vk,\ii\omega_n)$ is the gap function, and $\hat{\tau}_i$ are the Pauli matrices.
The Green's function $\hat{\mathbf{G}}(\vk,\ii\omega_n)$ is defined by the Dyson's equation
\begin{equation}
	\hat{\mathbf{G}}^{-1}(\vk,\ii\omega_n) = \hat{\mathbf{G}}_0^{-1}(\vk,\ii\omega_n) - \hat{\mathbf{\Sigma}}(\vk,\ii\omega_n), \label{eq:DysonEq}
\end{equation}
where $\hat{\mathbf{G}}_0(\vk,\ii\omega_n)$ is the bare Green's function,
\begin{equation}
	\hat{G}_{0\alpha}^{-1}(\vk,\ii\omega_n) =  \ii\omega_n\hat{\tau}_0 - \epsilon_\alpha(\vk)\hat{\tau}_3, \label{eq:bareGF}
\end{equation}
with $\epsilon_\alpha(\vk)$ being the single-electron energy counted from the Fermi level (chemical potential). Substituting equations~\eqref{eq:SelfEn} and~\eqref{eq:bareGF} into equation~\eqref{eq:DysonEq}, and inverting the result, we obtain the Green's function
\begin{align}
	&\hat{G}_{\alpha}(\vk,\ii\omega_n) = \det{[\hat{G}_\alpha^{-1}(\vk,\ii\omega_n)]}^{-1} \nonumber \\
	&\times \left[\ii\omega_nZ_\alpha(\vk,\ii\omega_n)\hat{\tau}_0 + \xi_\alpha(\vk,\ii\omega_n)\hat{\tau}_3 + \phi_{2\alpha}(\vk,\ii\omega_n)\hat{\tau}_2 \right], \label{eq:GF}
\end{align}
where $\xi_\alpha(\vk,\ii\omega_n) = \epsilon_\alpha(\vk) + \chi_\alpha(\vk,\ii\omega_n)$, and 
\begin{align}
	\det{[\hat{G}_\alpha^{-1}(\vk,\ii\omega_n)]} &= [\ii\omega_nZ_\alpha(\vk,\ii\omega_n)]^2 \nonumber \\ 
	&- \xi_\alpha^2(\vk,\ii\omega_n) - \phi_{2\alpha}^2(\vk,\ii\omega_n). \label{eq:detInvG}
\end{align}

To calculate the trace $\Tr\left[ \ln{\lbrace-\hat{\mathbf{G}}^{-1}(\vk,\ii\omega_n)\rbrace} \right]$, we employ very convenient property of the natural logarithm of a matrix,
\begin{equation}
	\Tr[\ln{\lbrace\hat{M}\rbrace}] = \ln{\lbrace\det{[\hat{M}]}\rbrace}.
\end{equation}
Since the full Green's function is the block-diagonal matrix in the combined space, we can immediately write the expression in the two-band case with $\alpha = \lbrace a,b \rbrace$ being the band index,
\begin{align}
	&\det{[-\hat{\mathbf{G}}^{-1}(\vk,\ii\omega_n)]} \nonumber \\ &=\det{[-\hat{G}_a^{-1}(\vk,\ii\omega_n)]}\det{[-\hat{G}_b^{-1}(\vk,\ii\omega_n)]} \nonumber\\
	&= (-1)^2\prod_{\alpha = a,b}\Bigl\{[\omega_nZ_\alpha(\vk,\ii\omega_n)]^2 \Bigr. \nonumber \\
	&+ \Bigl. \xi_\alpha^2(\vk,\ii\omega_n) + \phi_{2\alpha}^2(\vk,\ii\omega_n)\Bigr\}, \label{eq:MinusdetInvG}
\end{align}
and
\begin{align}
	&\ln{\lbrace\det{[-\hat{\mathbf{G}}^{-1}(\vk,\ii\omega_n)]}\rbrace} \nonumber\\
	&= \ln{\prod_{\alpha = a,b}\Bigl\{[\omega_nZ_\alpha(\vk,\ii\omega_n)]^2 \Bigr.} \nonumber \\
	&+ {\Bigl. \xi_\alpha^2(\vk,\ii\omega_n) + \phi_{2\alpha}^2(\vk,\ii\omega_n)\Bigr\}} \nonumber\\
	&= \sum_{\alpha = a,b}\ln{\Bigl\lbrace[\omega_nZ_\alpha(\vk,\ii\omega_n)]^2 \Bigr. } \nonumber \\
	&+ {\Bigl. \xi_\alpha^2(\vk,\ii\omega_n) + \phi_{2\alpha}^2(\vk,\ii\omega_n)\Bigr\rbrace}. \label{eq:lndet}
\end{align}

\subsection{$\xi$-integration}
Now, we should perform integration over momenta. We turn to a minimal model for iron-based materials that has two bands and lacks for momentum dependence still having the band indices. While it sacrifices quantitative precision, it preserves such qualitative peculiarities as the transition between $s_{\pm}$ and $s_{++}$ superconducting states~\cite{KorshunovUFN2016,EfremovKorshunov2011}. To simplify calculations, we assume that (implicit) dependence of functions $Z_{\alpha}$, $\chi_\alpha$ and $\phi_{2\alpha}$ on $\epsilon_\alpha$, and, consequently, on $\vk$ are negligible~\cite{allen}.

Thus, we assume that $\xi_\alpha(\vk,\ii\omega_n) = \epsilon_\alpha(\vk)$ and $\chi_\alpha$ now becomes just an additive to the chemical potential. Therefore,
\begin{equation}
	\hat{\Sigma}_{\alpha}(\ii\omega_n) = \ii\omega_n[1-Z_\alpha(\ii\omega_n)]\hat{\tau}_0 + \phi_{2\alpha}(\ii\omega_n)\hat{\tau}_2. \label{eq:SelfEn.dispLess}
\end{equation}
Although, generally, the self-energy matrix $\hat{\Sigma}_{\alpha}(\omega_n)$ contains terms proportional to all Pauli matrices, within the current model a term proportional to $\hat{\tau}_3$ is omitted under assumption of isotropic self-energy, and a term proportional to $\hat{\tau}_{1}$ is excluded due to the symmetry of the Eliashberg equations~\cite{FTT2024}.
 
Considering processes with energies near the Fermi level, we can employ the $\xi$-integration procedure
\begin{equation}
	\sum_{\vk}\leftrightarrow\int\frac{\mathrm{d}^3k}{(2\pi)^3} \rightarrow N_\alpha\int \mathrm{d}\xi_\alpha\int\frac{\mathrm{d}\theta}{4\pi}, \label{eq:xi.integration}
\end{equation}
where $N_\alpha$ is the density of states at the Fermi level per one electron spin projection in the $\alpha$-th band, 
the part $\int\mathrm{d}\theta/4\pi$ is the integration over directions that gives unity in the isotropic case. Therefore, we need take the following integral:
\begin{equation}
\int_{-\infty}^{\infty}\mathrm{d}\xi_\alpha\ln{\Bigl\lbrace[\omega_nZ_\alpha(\ii\omega_n)]^2 + \xi_\alpha^2  + \phi_{2\alpha}^2(\ii\omega_n)\Bigr\rbrace}. \label{eq:ln.to.integrate}
\end{equation}
The answer is
\begin{align}
	&-T\sum_{\omega_n,\vk}\Tr\left[ \ln{\lbrace-\hat{\mathbf{G}}^{-1}(\vk,\ii\omega_n)\rbrace} \right] \nonumber\\
	&= -T\sum_{\omega_n}\sum_{\alpha = a,b}\Bigl[N_\alpha I^{\alpha\infty}_{\SC} \Bigr. \nonumber \\ 
	&+ \Bigl. 2\pi N_\alpha\sqrt{[\omega_nZ_\alpha(\ii\omega_n)]^2 + \phi_{2\alpha}^2(\ii\omega_n)}\Bigr]. \label{eq:ln.integrated}
\end{align}
and it contains a diverging term $N_\alpha I^{\alpha\infty}_{\SC}$ that would be eliminated later.

In the second term in equation~(\ref{eq:LW}), it is convenient to perform integration over momenta~(\ref{eq:xi.integration}) before taking the trace. Under the aforementioned assumptions on $\vk$-dependency of functions $Z_{\alpha}$, $\chi_\alpha$, $\phi_{2\alpha}$, and the self-energy~(\ref{eq:SelfEn.dispLess}) on $\epsilon_\alpha$ the problem reduces to replacing the Green's function $\hat{\mathbf{G}}(\vk,\ii\omega_n)$ with the $\xi$-integrated one $\hat{\mathbf{g}}(\ii\omega_n)$:
\begin{align}
&-T\sum_{\omega_n,\vk}\Tr\left[\hat{\mathbf{\Sigma}}(\vk,\ii\omega_n)\hat{\mathbf{G}}(\vk,\ii\omega_n) \right] \nonumber\\
	&\rightarrow -T\sum_{\omega_n}\int \mathrm{d}\xi_\alpha\Tr\left[\hat{\mathbf{\Sigma}}(\ii\omega_n)\hat{\mathbf{G}}(\xi_\alpha(\vk),\ii\omega_n) \right] \nonumber\\
	&= -T\sum_{\omega_n}\Tr\left[\hat{\mathbf{\Sigma}}(\ii\omega_n)\hat{\mathbf{g}}(\ii\omega_n) \right], \label{eq:SigmG.xi.int}
\end{align}
where
\begin{align}
	\hat{g}_\alpha(\ii\omega_n) &= -\pi N_\alpha\frac{\ii\omega_nZ_\alpha(\ii\omega_n)\hat{\tau}_0 + \phi_{2\alpha}(\ii\omega_n)\hat{\tau_2}}{\sqrt{[\omega_nZ_\alpha(\ii\omega_n)]^2 + \phi_{2\alpha}^2(\ii\omega_n)}} \nonumber \\
	&= g_{0\alpha}\hat{\tau}_0 + g_{2\alpha}\hat{\tau}_2. \label{eq:g0g1g2}
\end{align}
Now, we introduce the notations $\ii\tomega_{\alpha n} = \ii\omega_nZ_\alpha(\ii\omega_n)$ and $\phi_{2\alpha n}^2 = \phi_{2\alpha}^2(\ii\omega_n)$. Calculating the trace, we obtain 
\begin{align} &-T\sum_{\omega_n}\Tr\left[\hat{\mathbf{\Sigma}}(\ii\omega_n)\hat{\mathbf{g}}(\ii\omega_n) \right] \nonumber\\
	&= -2\pi T\sum_{\omega_n}\sum_{\alpha = a,b} N_{\alpha}\left[ \frac{\omega_n\tomega_{\alpha n}}{\Qalphan} - \Qalphan\right], \label{eq:SigmG}
\end{align}
where $\Qalphan=\sqrt{\tomega_{\alpha n}^2 + \phi_{2\alpha n}^2}$.

\subsection{Eliashberg equations and self-energy}
Calculation of the self-energy and the thermodynamic potential should be done in the same approximation. To proceed with the impurity part $\Omega'_{\imp}$ and the superconducting part $\Omega'_{\SC}$ of the Luttinger-Ward functional~\eqref{eq:LW_f}, we first introduce procedure for calculating self-energies in un unconventional superconductor with nonmagnetic impurities~\cite{Ohashi2004,KorshunovUFN2016}.

The gap function is obtained by solving a system of the Eliashberg equations, which at the imaginary axis has the form
\begin{align}
	\ii\tomega_{\alpha n} = \ii\omega_n - \Sigma^{\SC}_{0\alpha}(\tomega_{\alpha n},\phi_{2\alpha n}) - \Sigma^{\imp}_{0\alpha}(\tomega_{\alpha n},\phi_{2\alpha n}), \label{eq:Eliashberg.omega} \\
	\phi_{2\alpha n} = \Sigma^{\SC}_{2\alpha}(\tomega_{\alpha n},\phi_{2\alpha n}) + \Sigma^{\imp}_{2\alpha}(\tomega_{\alpha n},\phi_{2\alpha n}), \label{eq:Eliashberg.phi}
\end{align}
where $\tilde{\omega}_{\alpha n}$ is the Matsubara  
frequency renormalized by self-energies connected with superconducting interaction $\Sigma^{\SC}_{0\alpha}(\omega_n)$ and impurity scattering $\Sigma^{\imp}_{0\alpha}(\omega_n)$; $\phi_{\alpha n} \equiv \phi_{\alpha}(\omega_n)$ is a superconducting order parameter defined by self-energies $\Sigma^{\SC}_{2\alpha}(\omega_n)$ and $\Sigma^{\imp}_{2\alpha}(\omega_n)$, and connected with the superconducting gap function $\Delta_{\alpha n} = \phi_{\alpha n} / Z_{\alpha n}$ by means of a renormalization factor $Z_{\alpha n} = \tomega_{\alpha n} / \omega_n$. Subscripts `$0$' and `$2$' in equations~\eqref{eq:Eliashberg.omega} and~\eqref{eq:Eliashberg.phi} denote corresponding Pauli matrices $\hat{\tau}_i$ within the Nambu space,
\begin{align}
	\hat{\Sigma}_{\alpha}(\tomega_{\alpha n},\phi_{2\alpha n}) &= \sum_{i=0,2}\Bigl[ \Sigma^{\SC}_{i\alpha}(\tomega_{\alpha n},\phi_{2\alpha n}) \Bigr.\nonumber\\
	&+ \Bigl. \Sigma^{\imp}_{i\alpha}(\tomega_{\alpha n},\phi_{2\alpha n})\Bigr]\hat{\tau}_i. \label{eq:selfEnergy.Nambu}
\end{align}
Explicitly, terms $\Sigma^{\SC}_{i\alpha}(\ii\omega_n)$ for the self-energy have the following form:
\begin{align}
	\Sigma^{\SC}_{0\alpha}(\tomega_{\alpha n},\phi_{2\alpha n}) = -\ii\pi T\sum_{\omega_{n'} \beta} \frac{\lambda^{Z}_{\alpha\beta}(\omega_n - \omega_{n'})\tomega_{\beta n'}}{\Qbetan}, \label{eq:Sigma.SC0} \\
	\Sigma^{\SC}_{2\alpha}(\tomega_{\alpha n},\phi_{2\alpha n}) = \pi T\sum_{\omega_{n'} \beta} \frac{\lambda^{\phi}_{\alpha\beta}(\omega_n - \omega_{n'})\phi_{\beta n'}}{\Qbetan}, \label{eq:Sigma.SC2}
\end{align}
where $\lambda^{Z,\phi}_{\alpha\beta}(\omega_n - \omega_{n'}) = 2\lambda^{Z,\phi}_{\alpha\beta}\int_0^\infty \mathrm{d}w\frac{w B(w)}{(\omega_n - \omega_{n'})^2 + w^2}$ is a coupling function, $\lambda^{\omega,\phi}_{\alpha\beta}$ is a coupling constant chosen for simplicity as $\lambda^{Z}_{\alpha\beta} = \left|\lambda^{\phi}_{\alpha\beta}\right| = \left|\lambda_{\alpha\beta}\right|$, $B(w)$ is a bosonic spectral function depending on real frequency $w$~\cite{KorshunovUFN2016}. The impurity part $\Sigma^{\imp}_{i\alpha}(\omega_n)$ of the self-energy is calculated within the $\mathcal{T}$-matrix approximation~\cite{Ohashi2004,KorshunovUFN2016} and has the form
\begin{align}
	&\Sigma^{\imp}_{0\alpha}(\tomega_{\alpha n},\phi_{2\alpha n}) = \frac{-\ii\Gamma_a}{2\Dimp}\left[\frac{\sigma(1-\eta^2)^2\tomega_{an}}{\Qan} \right.\nonumber\\
	&+ \left.(1-\sigma)\left( \frac{\eta^2N_a\tomega_{an}}{N_b\Qan} + \frac{\tomega_{bn}}{\Qbn}\right)\right], \label{eq:Sigma.imp0} \\
	&\Sigma^{\imp}_{2\alpha}(\tomega_{\alpha n},\phi_{2\alpha n}) = \frac{\Gamma_a}{2\Dimp}\left[\frac{\sigma(1-\eta^2)^2\phi_{an}}{\Qan} \right. \nonumber \\
	&+ \left. (1-\sigma)\left( \frac{\eta^2N_a\phi_{an}}{N_b\Qan} + \frac{\phi_{bn}}{\Qbn}\right)\right]. \label{eq:Sigma.imp2}
\end{align}
Here $\eta = v/u$ is a ratio between intraband ($v$) and interband ($u$) impurity potentials, $\Gamma_a$ is an impurity scattering rate
\begin{equation}
	\Gamma_a = \frac{2n_{\imp}\sigma}{\pi N_a} = 2n_{\imp}\pi N_b u^2(1-\sigma), \label{eq:Gamma_a}
\end{equation}
that depends on concentration of impurities $n_{\imp}$ and effective cross section
\begin{equation}
	\sigma = \frac{\pi^2N_aN_bu^2}{1 + \pi^2N_aN_bu^2}. \label{eq:sigma}
\end{equation}
Denominator $\Dimp$ is a short notation for the following expression:
\begin{equation}
	\Dimp = (1-\sigma)^2 + \sigma^2(1-\eta^2)^2 + \sigma(1-\sigma)\kappa_{\imp}, \label{eq:D_imp}
\end{equation}
where
\begin{equation}
	\kappa_{\imp} = \eta^2\frac{N_a^2 + N_b^2}{N_aN_b} + 2\frac{\tomega_{an}\tomega_{bn} + \phi_{2an}\phi_{2bn}}{\Qan\Qbn}. \label{eq:kappa.imp}
\end{equation}

\subsection{Impurity part of $\Omega$}
The impurity part $\Omega'_{\imp}$ of the Luttinger-Ward functional \eqref{eq:LW_f} is defined by the series of diagrams presented in Figure \ref{fig:imp.series}, with the corresponding expression:
\begin{figure}
    \centering
    \includegraphics[width=0.45\textwidth]{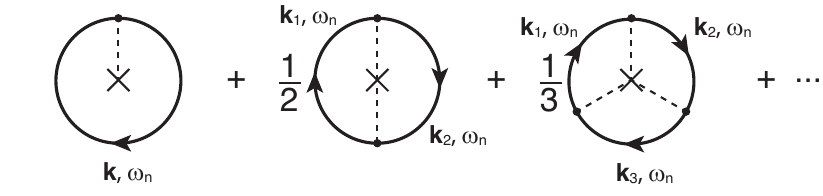}
    \caption{Diagrammatic series for the impurity term of the Luttinger-Ward functional $\Omega'_{\imp}$.}
    \label{fig:imp.series} 
\end{figure}
\begin{equation}
	\Omega'_{\imp}(T) = n_{\imp}T\sum_{\omega_n, \vk}\Tr\left[ \sum_{t=1}^{\infty}\frac{1}{t}\left\{\hat{\mathbf{U}}\hat{\mathbf{G}}(\vk, \ii\omega_n)\right\}^{t} \right], \label{eq:pos.series1}
\end{equation}
where $\hat{\mathbf{U}}$ is the impurity potential matrix. The series of diagrams in Figure \ref{fig:imp.series} is corresponding to diagrammatic series for the impurity self-energy within the $\mathcal{T}$-matrix approximation. In the case of the two-band system the impurity potential matrix has a simple form containing intraband and interband scattering potentials:
\begin{equation}
	\lbrace\hat{\mathbf{U}}\rbrace_{\alpha\beta} = \left[ u + (v - u)\delta_{\alpha\beta} \right]\otimes\hat{\tau}_3. \label{eq:Umatrix.red}
\end{equation}
In equation~\eqref{eq:pos.series1} we can perform integration over momenta~\eqref{eq:xi.integration} in each term of the series apart from its summation, which leads to substitution of the Green's function $\hat{\mathrm{G}}$ by the $\xi$-integrated one. The following summation of the series gives us the expression:
\begin{align}
	\Omega'_{\imp}(T) &= -n_{\imp}T\sum_{\omega_n}\Tr\left[\ln{\left\lbrace \hat{\mathbf{1}} - \hat{\mathbf{U}}\hat{\mathbf{g}}(\ii\omega_n)\right\rbrace}\right] \nonumber\\	
	&= -n_{\imp}T\sum_{\omega_n}\ln{\left\lbrace \det{\left[ \hat{\mathbf{1}} - \hat{\mathbf{U}}\hat{\mathbf{g}}(\ii\omega_n)\right]}\right\rbrace}. \label{eq:pos.series2}
\end{align}
where $\hat{\mathbf{1}}$ is the identity matrix in the band and Nambu spaces. In terms of parameters introduced by equations~\eqref{eq:Gamma_a}~-- \eqref{eq:kappa.imp} determinant in equation~\eqref{eq:pos.series2} is:
\begin{equation}
	\det{\left[ \hat{\mathbf{1}} - \hat{\mathbf{U}}\hat{\mathbf{g}}(\ii\omega_n)\right]} 
	= \Dimp/(1-\sigma)^2. \label{eq:det.imp}
\end{equation}
By substituting the equation \eqref{eq:det.imp} into \eqref{eq:pos.series2}, we obtain the expression
\begin{equation}
	\Omega'_{\imp} = -n_{\imp}T\sum_{\omega_n} \left[ \ln{\Dimp} - 2 \ln{(1-\sigma)} \right], \label{eq:Omega.imp}
\end{equation}
in which the second term in square brackets is changed from zero value in the Born limit 
($\sigma = 0$) to diverging one in the unitary limit 
($\sigma = 1$). As well as in the case of equation~\eqref{eq:ln.integrated}, this term will be eliminated later.

\subsection{Superconducting part of $\Omega$}
To calculate the superconducting part $\Omega'_{\SC}$ of the Luttinger-Ward functional, we treat combination of electron-phonon and spin-fluctuation interactions as an effective one with coupling constant $\hat{\mathbf{\lambda}}$. By analogy with the electron-phonon interaction, we assume that all diagrammatic expansion terms of the second and higher order with respect to this effective interaction do not make any significant contribution. Therefore, we can consider only the diagram, presented in Figure \ref{fig:sc.series}, leading to the following general expression:
\begin{figure}
    \centering
    \includegraphics[width=0.45\textwidth]{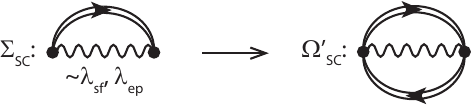}
    \caption{Self-energy diagram and the following from it diagram for the superconducting term of the Luttinger-Ward functional $\Omega'_{\SC}$. A wavy line represents an effective interaction, containing both spin-fluctuation and electron-phonon couplings.}
    \label{fig:sc.series}
\end{figure}
\begin{equation}
	\Omega'_{\SC} = \frac{T}{2}\sum_{\vk, \omega_n}\Tr{\left[ \hat{\mathbf{\Sigma}}_{\SC}(\vk,\ii\omega_n)\hat{\mathbf{G}}(\vk,\ii\omega_n) \right]}, \label{eq:SC.Omega}
\end{equation}
where $\hat{\mathbf{\Sigma}}_{\SC}(\vk,\ii\omega_n)$ is the self-energy corresponding to superconducting interaction only. Before calculating the trace of matrix product in~\eqref{eq:SC.Omega}, it is again convenient to perform the $\xi$-integration procedure that eliminates the $\vk$-dependency in the expression. Under the same assumptions which lead to equation~\eqref{eq:SelfEn.dispLess}, in the two-band case $\hat{\mathbf{\Sigma}}_{\SC}(\ii\omega_n)$ can be expressed from equations~\eqref{eq:Eliashberg.omega} and~\eqref{eq:Eliashberg.phi}. Since the latter system of equations, actually, is the self-consistent one, we cannot just use equations~\eqref{eq:Sigma.SC0} and~\eqref{eq:Sigma.SC2} to avoid double counting effect of impurities on Free energy. Rather, after $\ii\tomega_{\alpha n}$ and $\phi_{\alpha n}$ being calculated, we subtract the impurity part from the result and construct $\hat{\mathbf{\Sigma}}_{\SC}(\ii\omega_n)$:
\begin{align}
	\Sigma^{\SC}_{0\alpha}(\ii\omega_n) = \ii\tomega_{\alpha n} - \ii\omega_n +  \Sigma^{\imp}_{0\alpha}(\ii\omega_n), \label{eq:Subtract.omega} \\
	\Sigma^{\SC}_{2\alpha}(\ii\omega_n) = \phi_{2\alpha n} - \Sigma^{\imp}_{2\alpha}(\ii\omega_n), \label{eq:sbtrac.phi}
\end{align}
Having such a self-energy, we perform $\xi$-integration only for the Green's function. Taking the trace after doing so, we obtain the following expression for $\Omega'_{\SC}$:
\begin{align}
	\Omega'_{\SC} &= T\sum_{\omega_n}\sum_{\alpha = a,b}\pi N_\alpha\left[\frac{\omega_n\tomega_{\alpha n}}{\Qalphan} - \Qalphan\right] \nonumber\\
	& + T\sum_{\omega_n}
	\frac{\pi N_a\Gamma_a}{2\Dimp}
	\left[
	2\sigma\left(1-\eta^2\right)^2 + (1-\sigma)\kappa_{\imp} 
	\right]. \label{eq:trace.SC.ptime.ans}
\end{align}

\subsection{Thermodynamic potential in the superconducting and normal states}
By combining equations \eqref{eq:ln.integrated}, \eqref{eq:SigmG}, \eqref{eq:Omega.imp}, and \eqref{eq:trace.SC.ptime.ans}, we obtain the expression for the thermodynamic potential of a two-band superconductor with nonmagnetic impurities
\begin{align}
	&\Omega_{\mathrm{S}}(T) = -T\sum_{\omega_n}\sum_{\alpha = a,b}N_\alpha I^{\alpha\infty}_{\SC} \nonumber\\
	&- \pi T\sum_{\omega_n}\sum_{\alpha = a,b} N_{\alpha}\left[ \frac{\omega_n\tomega_{\alpha n}}{\Qalphan} + \Qalphan\right] \nonumber\\
	&+ T\sum_{\omega_n} \frac{\pi N_a\Gamma_a}{2\Dimp}
	\left[ 2\sigma\left(1-\eta^2\right)^2 + (1-\sigma)\kappa_{\imp} \right] \nonumber\\
	&- n_{\imp}T\sum_{\omega_n}\ln{\Dimp} + 2n_{\imp}T\sum_{\omega_n}\ln{(1-\sigma)}.\label{eq:LW2}
\end{align}

For the normal state, the Landau free energy of a two-band metal with nonmagnetic impurities has the same general expression as for the superconducting state,
\begin{align}
	&\Omega_{\mathrm{N}}(T) = -T\sum_{\omega_n,\vk}\Tr\left[ \ln{\lbrace-\hat{\mathbf{G}}^{-1}_{\mathrm{N}}(\vk,\ii\omega_n)\rbrace} \right. \nonumber \\
	&+ \left. \hat{\mathbf{\Sigma}}_{\mathrm{N}}(\vk,\ii\omega_n)\hat{\mathbf{G}}_{\mathrm{N}}(\vk,\ii\omega_n) \right] + \Omega'^{\mathrm{N}}(T), \label{eq:LW.N}
\end{align}
except that in all quantities here with subscript or superscript $\mathrm{N}$ the order parameter $\phi_{\alpha n}$ is set to zero. In the same vein as for the superconducting state, we obtain the following expression for $\Omega_{\mathrm{N}}(T)$:
\begin{align}
	&\Omega_{\mathrm{N}}(T) = -T\sum_{\omega_n}\sum_{\alpha = a,b} N_\alpha I^{\alpha\infty}_{\mathrm{NS}} \nonumber \\
	&- \pi T\sum_{\omega_n}\sum_{\alpha = a,b} N_{\alpha}\Bigl[ \left|\omega_n\right| + \left|\tomega_{\alpha n}^{\mathrm{N}}\right|\Bigr] \nonumber\\
	&+ T\sum_{\omega_n} \frac{\pi N_a\Gamma_a}{2\Dimp^{\mathrm{N}}}
	\left[ 2\sigma\left(1-\eta^2\right)^2 + (1-\sigma)\kappa_{\imp}^{\mathrm{N}} \right] \nonumber\\
	&- n_{\imp}T\sum_{\omega_n}\ln{\Dimp^{\mathrm{N}}} + 2n_{\imp}T\sum_{\omega_n}\ln{(1-\sigma)}. \label{eq:LW2N}
\end{align}
where $\kappa_{\imp}^{\mathrm{N}} = \eta^2\frac{N_a^2 + N_b^2}{N_aN_b} + 2$ and $\Dimp^{\mathrm{N}} = (1-\sigma)^2 + \sigma^2\left(1-\eta^2\right)^2 + \sigma(1-\sigma)\kappa_{\imp}^{\mathrm{N}}$.

Next, if we calculate difference between the thermodynamic potentials $\Delta\Omega = \Omega_{\mathrm{S}} - \Omega_{\mathrm{N}}$, the singular terms $I^{\alpha\infty}_{\mathrm{SC}}$ and $I^{\alpha\infty}_{\mathrm{NS}}$, as well as $2n_{\imp}T\sum_{\omega_n}\ln{(1-\sigma)}$, would chancel out each other. Finally, we arrive to the expression for the grand thermodynamic potential,
\begin{align}
	\Delta\Omega(T) &= - \pi T\sum_{\omega_n}\sum_{\alpha = a,b} N_{\alpha} \Bigl[ \frac{\omega_n\tomega_{\alpha n}}{\Qalphan} + \Qalphan \Bigr. \nonumber \\
	&- \Bigl. \left|\omega_n\right| - \left|\tomega_{\alpha n}^{\mathrm{N}}\right| \Bigr] \nonumber\\
	&+ \pi T N_a\Gamma_a\sum_{\omega_n}\left[ \frac{2\sigma\left(1-\eta^2\right)^2 + (1-\sigma)\kappa_{\imp}}{2\Dimp} \right. \nonumber \\
	&- \left. \frac{2\sigma\left(1-\eta^2\right)^2 + (1-\sigma)\kappa_{\imp}^{\mathrm{N}}}{2\Dimp^{\mathrm{N}}}
	\right] \nonumber\\
	&- n_{\imp}T\sum_{\omega_n}\ln{\frac{\Dimp}{\Dimp^\mathrm{N}}}. \label{eq:DLW2}
\end{align}

In the Born limit, expressing concentration of impurities $n_{\imp}$ through $\Gamma_a$, we should take the following limit for the last term:
\begin{align}
	&\lim_{\sigma\to0}\left\{\frac{\pi N_a\Gamma_a}{2\sigma}\ln{\frac{\Dimp}{\Dimp^{\mathrm{N}}}}\right\} \nonumber\\
	&= \pi T N_a\Gamma_a\sum_{\omega_n}\left[ \frac{\tomega_{an}\tomega_{bn}+\phi_{2an}\phi_{2bn}}{\Qan\Qbn} - 1 \right].
\end{align}
It turns out to be equal, up to sign, to the next to the last term in equation~(\ref{eq:DLW2}) in the Born limit:
\begin{align}
	&\lim_{\sigma\to0}\left\{\pi T N_a\Gamma_a\sum_{\omega_n}\left[ \frac{2\sigma(1-\eta^2)^2 + (1-\sigma)\kappa_{\imp}}{2\Dimp} \right.\right. \nonumber \\
	&- \left.\left. \frac{2\sigma(1-\eta^2)^2 + (1-\sigma)\kappa_{\imp}^{\mathrm{N}}}{2\Dimp^{\mathrm{N}}}\right] \right\} \nonumber\\
	&= \pi T N_a\Gamma_a\sum_{\omega_n}\left[ \frac{\tomega_{an}\tomega_{bn}+\phi_{2an}\phi_{2bn}}{\Qan\Qbn} - 1 \right]. \label{eq:limit.Born}
\end{align}
Thus, in the expression for $\Delta\Omega$ for the Born limit the last two terms chancels out, and $\Delta\Omega(T)$ becomes only implicitly dependent on impurity scattering through the renormalized Matsubara frequencies and gap function:
\begin{align}
	\Delta\Omega_{\mathrm{Born}}(T) &= - \pi T\sum_{\omega_n}\sum_{\alpha = a,b} N_{\alpha} \Bigl[ \frac{\omega_n\tomega_{\alpha n}}{\Qalphan} \Bigr. \nonumber \\
	&+ \Bigl. \Qalphan -\left|\omega_n\right| - \left|\tomega_{\alpha n}^{\mathrm{N}}\right| \Bigr]. \label{eq:DLW2Born}
\end{align}

\section{Results for iron-based materials} \label{sect:2.results}
For the impurity induced transition between $s_{\pm}$ and $s_{++}$ states to occur a sign of the averaged over bands coupling constant $\langle \lambda \rangle$ for the clean $s_{\pm}$ superconductor should be positive~\cite{EfremovKorshunov2011}. Thus below we use the following values for the coupling constants matrix elements: $\lbrace \lambda_{aa}, \lambda_{ab}, \lambda_{ba}, \lambda_{bb} \rbrace = \lbrace 3.0, -0.2, -0.1, 0.5 \rbrace$. In the clean limit it gives the superconducting state with critical temperature $T_{c0} = 40$~K. The densities of states $N_a = 1.0656$~eV$^{-1}$ and $N_b = 2N_a$ are chosen so that the value of total density of states $N = N_a + N_b$ being close to the one obtained within first-principle calculations~\cite{Ferber2010,Yin2011,Sadovskii2012}. We assume without loss of generality that the impurity scattering occurs in the interband channel only, $\eta = 0$, since nonzero intraband scattering potential is previously shown to have no influence on the superconducting state in the Born limit and for nonzero values of the cross-section $\sigma$ only shifts the transition point to higher values of $\Gamma_a$ \cite{ShestakovKorshunovSUST2018}.

\begin{figure}
    \centering
    \includegraphics[width=0.48\textwidth]{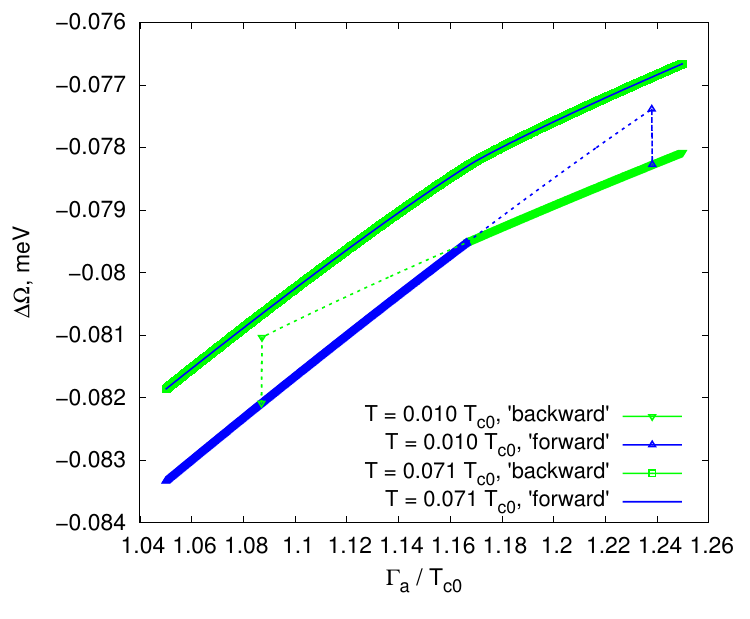}
    \caption{Dependence of $\Delta\Omega$ on the impurity scattering rate $\Gamma_a$ for two directions of obtaining solutions of the Eliashberg equations: `forward' is for adding nonmagnetic disorder starting from the clean limit, `backward' is for opposite direction of ``cleaning out'' the system. Dashed lines demonstrate energetically unfavorable branches, while solid lines with symbols correspond to  minimal energy.}
    \label{fig:free.Energy}
\end{figure}

In figure~\ref{fig:free.Energy}, the grand thermodynamic potential $\Delta\Omega(T=0.01T_{c0})$ in the Born limit $\sigma = 0$ is shown for two directions of the system evolution corresponding to the different types of solutions: `forward' is for increasing amount of disorder in the superconductor starting from the clean limit, `backward' is for ``cleaning out'' the system from initially disordered state with $\Gamma_a = 6T_{c0}$ to the clean limit with $\Gamma_a=0$.
Comparing values of thermodynamic potential, i.e. Landau free energy, for these solutions, we choose energetically favorable one. Such a solutions are shown by the solid lines in figure~\ref{fig:free.Energy}. Switching between solutions produce a kink on the lowest curve. In the figure, $\Delta\Omega$ is also plotted for $T=0.071T_{c0}$ where the $s_{\pm} \to s_{++}$ transition is smooth and the Eliashberg equations have only a single set of solutions. At this temperature, thermodynamic potential has no peculiarities and solutions are equal for both directions, `forward' and `backward'.

\section{Conclusion} \label{sect:5.concl}
We derived the expression for the grand thermodynamic potential $\Delta\Omega = \Omega_{\mathrm{S}} - \Omega_{\mathrm{N}}$ using the Luttinger-Ward approach for the two-band model of iron-based superconductors with nonmagnetic impurities. In axes ($T$, $\Gamma_a$) there is an area where two solutions of the Eliashberg equations exist. They have different values of thermodynamic potential and can be obtained considering different directions of the system evolution with respect to adding disorder. Choosing solutions with the lowest $\Delta\Omega$ values leads to kink in the $\Delta\Omega(\Gamma_a)$ dependence. The kink, along with a hysteresis in the Eliashberg equations solutions~\cite{FTT2025}, indicates that the abrupt $s_{\pm} \to s_{++}$ transition at low temperature $T < 0.1$~$T_{c0}$ and in the vicinity of the Born limit is the first order phase transition.

\bmhead{Acknowledgements}

The study was supported by the Russian Science Foundation grant N25-22-20043, \url{https://rscf.ru/project/25-22-20043/}, grant of the Krasnoyarsk Regional Science Foundation.


\bibliography{Shestakov-JSNM-FPS2025_arxiv}


\begin{thebibliography}{25}
\ifx \bisbn   \undefined \def \bisbn  #1{ISBN #1}\fi
\ifx \binits  \undefined \def \binits#1{#1}\fi
\ifx \bauthor  \undefined \def \bauthor#1{#1}\fi
\ifx \batitle  \undefined \def \batitle#1{#1}\fi
\ifx \bjtitle  \undefined \def \bjtitle#1{#1}\fi
\ifx \bvolume  \undefined \def \bvolume#1{\textbf{#1}}\fi
\ifx \byear  \undefined \def \byear#1{#1}\fi
\ifx \bissue  \undefined \def \bissue#1{#1}\fi
\ifx \bfpage  \undefined \def \bfpage#1{#1}\fi
\ifx \blpage  \undefined \def \blpage #1{#1}\fi
\ifx \burl  \undefined \def \burl#1{\textsf{#1}}\fi
\ifx \doiurl  \undefined \def \doiurl#1{\url{https://doi.org/#1}}\fi
\ifx \betal  \undefined \def \betal{\textit{et al.}}\fi
\ifx \binstitute  \undefined \def \binstitute#1{#1}\fi
\ifx \binstitutionaled  \undefined \def \binstitutionaled#1{#1}\fi
\ifx \bctitle  \undefined \def \bctitle#1{#1}\fi
\ifx \beditor  \undefined \def \beditor#1{#1}\fi
\ifx \bpublisher  \undefined \def \bpublisher#1{#1}\fi
\ifx \bbtitle  \undefined \def \bbtitle#1{#1}\fi
\ifx \bedition  \undefined \def \bedition#1{#1}\fi
\ifx \bseriesno  \undefined \def \bseriesno#1{#1}\fi
\ifx \blocation  \undefined \def \blocation#1{#1}\fi
\ifx \bsertitle  \undefined \def \bsertitle#1{#1}\fi
\ifx \bsnm \undefined \def \bsnm#1{#1}\fi
\ifx \bsuffix \undefined \def \bsuffix#1{#1}\fi
\ifx \bparticle \undefined \def \bparticle#1{#1}\fi
\ifx \barticle \undefined \def \barticle#1{#1}\fi
\bibcommenthead
\ifx \bconfdate \undefined \def \bconfdate #1{#1}\fi
\ifx \botherref \undefined \def \botherref #1{#1}\fi
\ifx \url \undefined \def \url#1{\textsf{#1}}\fi
\ifx \bchapter \undefined \def \bchapter#1{#1}\fi
\ifx \bbook \undefined \def \bbook#1{#1}\fi
\ifx \bcomment \undefined \def \bcomment#1{#1}\fi
\ifx \oauthor \undefined \def \oauthor#1{#1}\fi
\ifx \citeauthoryear \undefined \def \citeauthoryear#1{#1}\fi
\ifx \endbibitem  \undefined \def \endbibitem {}\fi
\ifx \bconflocation  \undefined \def \bconflocation#1{#1}\fi
\ifx \arxivurl  \undefined \def \arxivurl#1{\textsf{#1}}\fi
\csname PreBibitemsHook\endcsname

\bibitem[\protect\citeauthoryear{Sadovskii}{2008}]{SadovskiiReview2008}
\begin{barticle}
\bauthor{\bsnm{Sadovskii}, \binits{M.V.}}:
\batitle{High-temperature superconductivity in iron-based layered compounds}.
\bjtitle{Phys. Usp.}
\bvolume{51}(\bissue{12}),
\bfpage{1201}--\blpage{1227}
(\byear{2008})
\doiurl{10.1070/PU2008v051n12ABEH006820}
\end{barticle}
\endbibitem

\bibitem[\protect\citeauthoryear{Stewart}{2011}]{StewartReview}
\begin{barticle}
\bauthor{\bsnm{Stewart}, \binits{G.R.}}:
\batitle{Superconductivity in iron compounds}.
\bjtitle{Rev. Mod. Phys.}
\bvolume{83},
\bfpage{1589}--\blpage{1652}
(\byear{2011})
\doiurl{10.1103/RevModPhys.83.1589}
\end{barticle}
\endbibitem

\bibitem[\protect\citeauthoryear{Hirschfeld
  et~al.}{2011}]{HirschfeldKorshunov2011}
\begin{barticle}
\bauthor{\bsnm{Hirschfeld}, \binits{P.J.}},
\bauthor{\bsnm{Korshunov}, \binits{M.M.}},
\bauthor{\bsnm{Mazin}, \binits{I.I.}}:
\batitle{Gap symmetry and structure of fe-based superconductors}.
\bjtitle{Reports on Progress in Physics}
\bvolume{74}(\bissue{12}),
\bfpage{124508}
(\byear{2011})
\end{barticle}
\endbibitem

\bibitem[\protect\citeauthoryear{Korshunov}{2014}]{Korshunov2014eng}
\begin{barticle}
\bauthor{\bsnm{Korshunov}, \binits{M.M.}}:
\batitle{Superconducting state in iron-based materials and spin-fluctuation
  pairing theory}.
\bjtitle{Physics-Uspekhi}
\bvolume{57}(\bissue{8}),
\bfpage{813}--\blpage{819}
(\byear{2014})
\doiurl{10.3367/UFNe.0184.201408h.0882}
\end{barticle}
\endbibitem

\bibitem[\protect\citeauthoryear{Prozorov et~al.}{2014}]{Prozorov2014}
\begin{barticle}
\bauthor{\bsnm{Prozorov}, \binits{R.}},
\bauthor{\bsnm{Ko\'{n}czykowski}, \binits{M.}},
\bauthor{\bsnm{Tanatar}, \binits{M.A.}},
\bauthor{\bsnm{Thaler}, \binits{A.}},
\bauthor{\bsnm{Bud'ko}, \binits{S.L.}},
\bauthor{\bsnm{Canfield}, \binits{P.C.}},
\bauthor{\bsnm{Mishra}, \binits{V.}},
\bauthor{\bsnm{Hirschfeld}, \binits{P.J.}}:
\batitle{Effect of electron irradiation on superconductivity in single crystals
  of
  $\mathrm{Ba}({\mathrm{fe}}_{1\ensuremath{-}x}{\mathrm{ru}}_{x}{)}_{2}{\mathrm{as}}_{2}$
  ($x=0.24$)}.
\bjtitle{Phys. Rev. X}
\bvolume{4},
\bfpage{041032}
(\byear{2014})
\doiurl{10.1103/PhysRevX.4.041032}
\end{barticle}
\endbibitem

\bibitem[\protect\citeauthoryear{Inosov}{2016}]{Inosov2016}
\begin{barticle}
\bauthor{\bsnm{Inosov}, \binits{D.S.}}:
\batitle{Spin fluctuations in iron pnictides and chalcogenides: From
  antiferromagnetism to superconductivity}.
\bjtitle{Comptes Rendus Physique}
\bvolume{17}(\bissue{1-2}),
\bfpage{60}--\blpage{89}
(\byear{2016})
\doiurl{10.1016/j.crhy.2015.03.001}
\end{barticle}
\endbibitem

\bibitem[\protect\citeauthoryear{Schilling et~al.}{2016}]{Schilling2016}
\begin{barticle}
\bauthor{\bsnm{Schilling}, \binits{M.B.}},
\bauthor{\bsnm{Baumgartner}, \binits{A.}},
\bauthor{\bsnm{Gorshunov}, \binits{B.}},
\bauthor{\bsnm{Zhukova}, \binits{E.S.}},
\bauthor{\bsnm{Dravin}, \binits{V.A.}},
\bauthor{\bsnm{Mitsen}, \binits{K.V.}},
\bauthor{\bsnm{Efremov}, \binits{D.V.}},
\bauthor{\bsnm{Dolgov}, \binits{O.V.}},
\bauthor{\bsnm{Iida}, \binits{K.}},
\bauthor{\bsnm{Dressel}, \binits{M.}},
\bauthor{\bsnm{Zapf}, \binits{S.}}:
\batitle{Tracing the ${s}_{\ifmmode\pm\else\textpm\fi{}}$ symmetry in iron
  pnictides by controlled disorder}.
\bjtitle{Phys. Rev. B}
\bvolume{93},
\bfpage{174515}
(\byear{2016})
\doiurl{10.1103/PhysRevB.93.174515}
\end{barticle}
\endbibitem

\bibitem[\protect\citeauthoryear{Korshunov}{2018}]{Korshunov2018}
\begin{barticle}
\bauthor{\bsnm{Korshunov}, \binits{M.M.}}:
\batitle{Effect of gap anisotropy on the spin resonance peak in the
  superconducting state of iron-based materials}.
\bjtitle{Phys. Rev. B}
\bvolume{98},
\bfpage{104510}
(\byear{2018})
\doiurl{10.1103/PhysRevB.98.104510}
\end{barticle}
\endbibitem

\bibitem[\protect\citeauthoryear{Ghigo et~al.}{2018}]{Ghigo2018}
\begin{barticle}
\bauthor{\bsnm{Ghigo}, \binits{G.}},
\bauthor{\bsnm{Torsello}, \binits{D.}},
\bauthor{\bsnm{Ummarino}, \binits{G.A.}},
\bauthor{\bsnm{Gozzelino}, \binits{L.}},
\bauthor{\bsnm{Tanatar}, \binits{M.A.}},
\bauthor{\bsnm{Prozorov}, \binits{R.}},
\bauthor{\bsnm{Canfield}, \binits{P.C.}}:
\batitle{Disorder-driven transition from ${s}_{\ifmmode\pm\else\textpm\fi{}}$
  to ${s}_{++}$ superconducting order parameter in proton irradiated
  $\mathrm{Ba}({\mathrm{fe}}_{1\ensuremath{-}x}{\mathrm{rh}}_{x}{)}_{2}{\mathrm{as}}_{2}$
  single crystals}.
\bjtitle{Phys. Rev. Lett.}
\bvolume{121},
\bfpage{107001}
(\byear{2018})
\doiurl{10.1103/PhysRevLett.121.107001}
\end{barticle}
\endbibitem

\bibitem[\protect\citeauthoryear{Korshunov
  et~al.}{2022}]{KorshunovKuzmichev2022}
\begin{barticle}
\bauthor{\bsnm{Korshunov}, \binits{M.M.}},
\bauthor{\bsnm{Kuzmichev}, \binits{S.A.}},
\bauthor{\bsnm{Kuzmicheva}, \binits{T.E.}}:
\batitle{Direct observation of the spin exciton in andreev spectroscopy of
  iron-based superconductors}.
\bjtitle{Materials}
\bvolume{15}(\bissue{17}),
\bfpage{6120}
(\byear{2022})
\doiurl{10.3390/ma15176120}
\end{barticle}
\endbibitem

\bibitem[\protect\citeauthoryear{Golubov and Mazin}{1997}]{Golubov1997}
\begin{barticle}
\bauthor{\bsnm{Golubov}, \binits{A.A.}},
\bauthor{\bsnm{Mazin}, \binits{I.I.}}:
\batitle{Effect of magnetic and nonmagnetic impurities on highly anisotropic
  superconductivity}.
\bjtitle{Phys. Rev. B}
\bvolume{55},
\bfpage{15146}--\blpage{15152}
(\byear{1997})
\doiurl{10.1103/PhysRevB.55.15146}
\end{barticle}
\endbibitem

\bibitem[\protect\citeauthoryear{Ohashi}{2004}]{Ohashi2004}
\begin{barticle}
\bauthor{\bsnm{Ohashi}, \binits{Y.}}:
\batitle{Effects of interband impurity scattering on superconducting density of
  states in a two-band superconductor}.
\bjtitle{Physica C: Superconductivity}
\bvolume{412–414 Part 1}(\bissue{0}),
\bfpage{41}--\blpage{45}
(\byear{2004})
\doiurl{10.1016/j.physc.2003.11.062} .
\bcomment{Proceedings of the 16th International Symposium on Superconductivity
  (ISS 2003). Advances in Superconductivity XVI. Part I}
\end{barticle}
\endbibitem

\bibitem[\protect\citeauthoryear{Korshunov et~al.}{2016}]{KorshunovUFN2016}
\begin{barticle}
\bauthor{\bsnm{Korshunov}, \binits{M.M.}},
\bauthor{\bsnm{Togushova}, \binits{Y.N.}},
\bauthor{\bsnm{Dolgov}, \binits{O.V.}}:
\batitle{Impurities in multiband superconductors}.
\bjtitle{Physics-Uspekhi}
\bvolume{59}(\bissue{12}),
\bfpage{1211}
(\byear{2016})
\doiurl{10.3367/UFNe.2016.07.037863}
\end{barticle}
\endbibitem

\bibitem[\protect\citeauthoryear{Efremov et~al.}{2011}]{EfremovKorshunov2011}
\begin{barticle}
\bauthor{\bsnm{Efremov}, \binits{D.V.}},
\bauthor{\bsnm{Korshunov}, \binits{M.M.}},
\bauthor{\bsnm{Dolgov}, \binits{O.V.}},
\bauthor{\bsnm{Golubov}, \binits{A.A.}},
\bauthor{\bsnm{Hirschfeld}, \binits{P.J.}}:
\batitle{Disorder-induced transition between $s_\pm$ and ${s}_{++}$ states in
  two-band superconductors}.
\bjtitle{Phys. Rev. B}
\bvolume{84},
\bfpage{180512}
(\byear{2011})
\doiurl{10.1103/PhysRevB.84.180512}
\end{barticle}
\endbibitem

\bibitem[\protect\citeauthoryear{Shestakov
  et~al.}{2018a}]{ShestakovKorshunovSymmetry2018}
\begin{barticle}
\bauthor{\bsnm{Shestakov}, \binits{V.A.}},
\bauthor{\bsnm{Korshunov}, \binits{M.M.}},
\bauthor{\bsnm{Dolgov}, \binits{O.V.}}:
\batitle{Temperature-dependent $s_{\pm} \leftrightarrow s_{++}$ transitions in
  the multiband model for fe-based superconductors with impurities}.
\bjtitle{Symmetry}
\bvolume{10}(\bissue{8}),
\bfpage{323}
(\byear{2018})
\doiurl{10.3390/sym10080323}
\end{barticle}
\endbibitem

\bibitem[\protect\citeauthoryear{Shestakov
  et~al.}{2018b}]{ShestakovKorshunovSUST2018}
\begin{barticle}
\bauthor{\bsnm{Shestakov}, \binits{V.A.}},
\bauthor{\bsnm{Korshunov}, \binits{M.M.}},
\bauthor{\bsnm{Togushova}, \binits{Y.N.}},
\bauthor{\bsnm{Efremov}, \binits{D.V.}},
\bauthor{\bsnm{Dolgov}, \binits{O.V.}}:
\batitle{Details of the disorder-induced transition between $s_{\pm}$ and
  $s_{++}$ states in the two-band model for fe-based superconductors}.
\bjtitle{Superconductor Science and Technology}
\bvolume{31}(\bissue{3}),
\bfpage{034001}
(\byear{2018})
\end{barticle}
\endbibitem

\bibitem[\protect\citeauthoryear{Shestakov and
  Korshunov}{2025a}]{Shestakov_Korshunov_SUST2025}
\begin{botherref}
\oauthor{\bsnm{Shestakov}, \binits{V.A.}},
\oauthor{\bsnm{Korshunov}, \binits{M.M.}}:
Thermodynamics of $s_{\pm}$-to-$s_{++}$ transition in iron pnictides in the
  vicinity of the born limit
\textbf{38}(5),
055002
(2025)
\doiurl{10.1088/1361-6668/adc8dd}
\end{botherref}
\endbibitem

\bibitem[\protect\citeauthoryear{Shestakov and Korshunov}{2025b}]{FTT2025}
\begin{barticle}
\bauthor{\bsnm{Shestakov}, \binits{V.A.}},
\bauthor{\bsnm{Korshunov}, \binits{M.M.}}:
\batitle{On the abrupt change in the superconducting order parameter in the
  vicinity of the $s_{\pm} \to s_{++}$ transition in the born limit}.
\bjtitle{Rus. Physics of the Solid State}
\bvolume{67}(\bissue{7}),
\bfpage{1262}
(\byear{2025})
\doiurl{10.61011/FTT.2025.07.61183.35HH-25}
\end{barticle}
\endbibitem

\bibitem[\protect\citeauthoryear{Luttinger and Ward}{1960}]{LW1960II}
\begin{barticle}
\bauthor{\bsnm{Luttinger}, \binits{J.M.}},
\bauthor{\bsnm{Ward}, \binits{J.C.}}:
\batitle{Ground-state energy of a many-fermion system. ii}.
\bjtitle{Phys. Rev.}
\bvolume{118},
\bfpage{1417}--\blpage{1427}
(\byear{1960})
\doiurl{10.1103/PhysRev.118.1417}
\end{barticle}
\endbibitem

\bibitem[\protect\citeauthoryear{Luttinger}{1960}]{Luttinger1960}
\begin{barticle}
\bauthor{\bsnm{Luttinger}, \binits{J.M.}}:
\batitle{Fermi surface and some simple equilibrium properties of a system of
  interacting fermions}.
\bjtitle{Phys. Rev.}
\bvolume{119},
\bfpage{1153}--\blpage{1163}
(\byear{1960})
\doiurl{10.1103/PhysRev.119.1153}
\end{barticle}
\endbibitem

\bibitem[\protect\citeauthoryear{Allen and Mitrovic}{1982}]{allen}
\begin{bchapter}
\bauthor{\bsnm{Allen}, \binits{P.B.}},
\bauthor{\bsnm{Mitrovic}, \binits{B.}}:
\bctitle{Theory of superconducting $t_c$}.
In: \beditor{\bsnm{Erenreich}, \binits{H.}},
\beditor{\bsnm{Zeitz}, \binits{F.}},
\beditor{\bsnm{Turnbull}, \binits{D.}} (eds.)
\bbtitle{Solid State Physics: Advances in Research and Applications}
vol. \bseriesno{37},
pp. \bfpage{1}--\blpage{92}.
\bpublisher{Academic},
\blocation{New York}
(\byear{1982})
\end{bchapter}
\endbibitem

\bibitem[\protect\citeauthoryear{Shestakov and Korshunov}{2024}]{FTT2024}
\begin{barticle}
\bauthor{\bsnm{Shestakov}, \binits{V.A.}},
\bauthor{\bsnm{Korshunov}, \binits{M.M.}}:
\batitle{A relative phase between components of $s_\pm$ superconducting order
  parameter within a two-band model with impurities}.
\bjtitle{Rus. Physics of the Solid State}
\bvolume{66}(\bissue{8}),
\bfpage{1258}
(\byear{2024})
\doiurl{10.61011/FTT.2024.08.58585.51HH}
\end{barticle}
\endbibitem

\bibitem[\protect\citeauthoryear{Ferber et~al.}{2010}]{Ferber2010}
\begin{barticle}
\bauthor{\bsnm{Ferber}, \binits{J.}},
\bauthor{\bsnm{Zhang}, \binits{Y.-Z.}},
\bauthor{\bsnm{Jeschke}, \binits{H.O.}},
\bauthor{\bsnm{Valent\'{\i}}, \binits{R.}}:
\batitle{Analysis of spin-density wave conductivity spectra of iron pnictides
  in the framework of density functional theory}.
\bjtitle{Phys. Rev. B}
\bvolume{82},
\bfpage{165102}
(\byear{2010})
\doiurl{10.1103/PhysRevB.82.165102}
\end{barticle}
\endbibitem

\bibitem[\protect\citeauthoryear{Yin et~al.}{2011}]{Yin2011}
\begin{barticle}
\bauthor{\bsnm{Yin}, \binits{Z.P.}},
\bauthor{\bsnm{Haule}, \binits{K.}},
\bauthor{\bsnm{Kotliar}, \binits{G.}}:
\batitle{Magnetism and charge dynamics in iron pnictides}.
\bjtitle{Nature Physics}
\bvolume{7},
\bfpage{294}
(\byear{2011})
\doiurl{10.1038/nphys1923}
\end{barticle}
\endbibitem

\bibitem[\protect\citeauthoryear{Sadovskii et~al.}{2012}]{Sadovskii2012}
\begin{barticle}
\bauthor{\bsnm{Sadovskii}, \binits{M.V.}},
\bauthor{\bsnm{Kuchinskii}, \binits{E.Z.}},
\bauthor{\bsnm{Nekrasov}, \binits{I.A.}}:
\batitle{Iron based superconductors: Pnictides versus chalcogenides}.
\bjtitle{Journal of Magnetism and Magnetic Materials}
\bvolume{324}(\bissue{21}),
\bfpage{3481}--\blpage{3486}
(\byear{2012})
\doiurl{10.1016/j.jmmm.2012.02.071} .
\bcomment{Fifth Moscow international symposium on magnetism}
\end{barticle}
\endbibitem

\end{thebibliography}

\end{document}